\documentclass{eas}
\usepackage{graphicx}
\usepackage{url}
\usepackage[authoryear]{natbib}
\usepackage[fleqn]{amsmath}
\usepackage[]{amsfonts}%
\usepackage{xcolor}
\usepackage{amssymb}
\def\mea{{\rm M_\oplus}}
\def\mjup{{\rm M_{Jup}}}
\def\mp{M_{\rm p}}
\def\msol{{\rm M_\odot}}
\def\rsol{{\rm R_\odot}}
\begin{document}

%%-----------------------------
%%      the top matter
%%-----------------------------
\title{Evolution of exoplanets and their parent stars} 
\author{Tristan Guillot}\address{Laboratoire Lagrange, UMR 7293, Universit\'e de Nice-Sophia Antipolis, CNRS, Observatoire de la C\^ote d'Azur, 
    06304 Nice Cedex 04, France}
\author{Douglas N.C. Lin}\address{UCO/Lick Observatory, University of California, Santa Cruz, CA 95064, USA}
\author{Pierre Morel$^1$}
\author{Mathieu Havel$^1$}
\author{Vivien Parmentier$^1$}
\begin{abstract}
Studying exoplanets with their parent stars is crucial to understand their population, formation and history. We review some of the key questions regarding their evolution with particular emphasis on giant gaseous exoplanets orbiting close to solar-type stars. 
For masses above that of Saturn, transiting exoplanets have large radii indicative of the presence of a massive hydrogen-helium envelope. Theoretical models show that this envelope progressively cools and contracts with a rate of energy loss inversely proportional to the planetary age. The combined measurement of planetary mass, radius and a constraint on the (stellar) age enables a global determination of the amount of heavy elements present in the planet interior. The comparison with stellar metallicity shows a correlation between the two, indicating that accretion played a crucial role in the formation of planets. The dynamical evolution of exoplanets also depends on the properties of the central star. We show that the lack of massive giant planets and brown dwarfs in close orbit around G-dwarfs and their presence around F-dwarfs are probably tied to the different properties of dissipation in the stellar interiors. Both the evolution and the composition of stars and planets are intimately linked.  
\end{abstract}
\maketitle
%%-----------------------------
%%      your text
%%-----------------------------
\section{Introduction}

The discovery of 51\,Peg\,b \citep{Mayor+Queloz1995} heralded the birth of a new field: exoplanetology. The study of exoplanets is important in itself, but they also offer us a new window to study their parent stars, the interactions between stars and planets and the processes that led to their formation. 

This paper stems from a presentation given in October 2013 in Roscoff, France, at the school ``The ages of stars''. We present a few key elements based on the authors' work to understand the physical and dynamical evolution of exoplanets and their parent stars. This is by no means a proper review of this extremely rich field. 

We first examine the physical mechanisms that govern the evolution of fluid planets, brown dwarfs and stars. In Section~3, we show that the global composition of exoplanets may be derived from evolution models and linked to the composition of their parent star. In Section~4, we discuss how star-planet interactions shape the population of close-in exoplanets and brown dwarfs, leading some to be swallowed by their star.

\section{Thermal evolution of exoplanets}

The wide range of exoplanets discovered include objects from a fraction of an Earth-mass to giant planets many times the mass of our Jupiter. It also includes objects orbiting only a few stellar radii away from their parent stars to objects at orbital distances beyond 100\,AU. We focus on objects for which both a mass and a radius may be determined, corresponding to the ones discovered transiting in front of their parent star. We also choose to restrict ourselves to giant planets and brown dwarfs, i.e. fluid objects mostly made of hydrogen and helium. Because of their large masses (more than about 100 times the mass of the Earth), these objects inherit a large amount of internal energy from their gravitational contraction. Because of a non-negligible thermal expansion coefficient, the progressive loss of this initial energy results in both a cooling and a contraction of the planets. It can indeed be shown that the energy loss per unit time (i.e. the {\it intrinsic luminosity}) follows a modified Kelvin-Helmholtz relation:
 \begin{equation}
L\approx \eta {GM^2\over R\tau},
\label{eq:lapprox}
\end{equation}
where $\tau$ is the age, and $\eta$ is a factor that hides most of the
complex physics. In the approximation that Coulomb and exchange
terms can be neglected, $\eta\approx\theta/(\theta +1)$ where $\theta$ is the electron degeneracy factor. The poor
compressibility of giant planets in their mature evolution stages
imply that $\eta\ll 1$ ($\eta\sim 0.03$ for Jupiter): the luminosity
is not obtained from the entire gravitational potential, but from the
much more limited reservoir constituted by the thermal internal
energy \citep{Guillot2005}. The evolution of giant planets and brown dwarfs is thus akin to the pre-main sequence evolution of stars \citep[see also][]{Burrows+97,Chabrier+Baraffe2000}. 

Giant planets thus gradually cool and contract but with a timescale that depends also on the amount of irradiation from their parent star \citep[e.g.][]{Hubbard1977}. Giant planets and brown dwarfs are thought to be mostly convective but planets very close to their star (such as 51 Peg b) develop an outer radiative zone that progressively extends to the deeper levels \citep{Guillot+96,GS02}. This radiative zone and the atmosphere to which it is linked \citep[see][]{Parmentier+Guillot2014} govern the rate of cooling and contraction. 

Another factor affecting the sizes of exoplanets is whether they contain more or less heavy elements\footnote{Here, anything other than hydrogen and helium is dubbed ``heavy element''.}: Everything else being equal, planets with a higher fraction of heavy elements tend to be smaller.

\begin{figure}
\includegraphics[width=\hsize,angle=0]{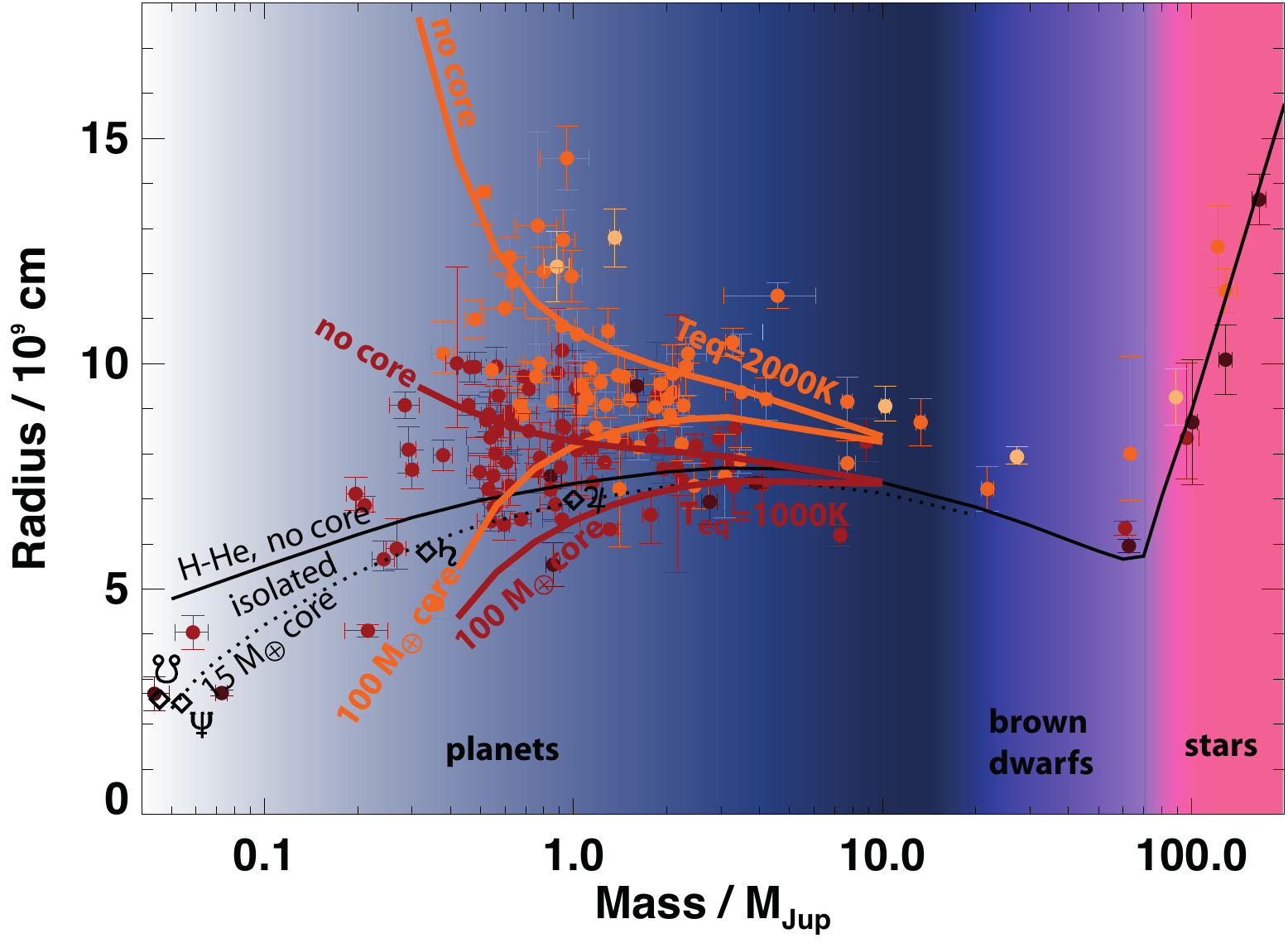}
\caption{Theoretical and observed mass-radius relations. The black
  line is applicable to the evolution of solar composition planets,
  brown dwarfs and stars, when isolated or nearly isolated (as
  Jupiter, Saturn, Uranus and Neptune, defined by diamonds and their
  respective symbols), after 5 Ga of evolution. The dotted line shows
  the effect of a $15\mea$ core on the mass-radius relation. Orange
  and yellow curves represent the mass-radius relations for heavily
  irradiated planets with equilibrium temperatures of 1000 and
  2000\,K, respectively, and assuming that 0.5\% of the incoming
  stellar luminosity is dissipated at the center. For each irradiation level, two cases
  are considered: a solar-composition planet with no core (top curve),
  and one with a $100\mea$ central core (bottom curve). Circles with error bars correspond to known planets, brown dwarfs and low-mass stars, color-coded as a function of their equilibrium temperature(below 750, 1500, 2250\,K and above 2250\,K, respectively, from darkest to lightest). [From \citep{Guillot+Gautier2014}]}
\label{fig:exo_M_R}
\end{figure}

Figure~\ref{fig:exo_M_R} compares a mass-radius diagram of known exoplanets, brown dwarfs and low-mass stars against some theoretical relations. After about 5 Ga of evolution, the mass-radius curve for non-irradiated hydrogen-helium planets has a maximum for a mass of $4\,\rm M_{Jup}$ and a radius slightly above that of Jupiter. For planets of larger masses, degeneracy effects in the equation of state begin to dominate, thermal effects become less important and cannot oppose as efficiently the increased compression: the radius decreases with mass until in the stellar domain nuclear reactions set in. However, for heavily-irradiated hydrogen-helium planets (i.e. for ``hot Jupiters''), the radius tends to decrease with mass: for smaller masses the warm atmosphere is less bounded to the planet because of the weaker gravity \citep[see][]{Guillot2005}. The presence of a massive core removes that effect. The variety of irradiation level and compositions is responsible for an ensemble of known giant exoplanet that is ``trumpet-shaped'' in that diagram.

\section{A link between stellar and exoplanetary compositions}

The possibility to measure planetary masses, radii, irradiation levels and to constrain the age through the stellar age thus offers the possibility to determine the global compositions of giant exoplanets. However, a problem remains: a significant fraction of hot Jupiters is found to have radii above that theoretically predicted for hydrogen-helium planets \citep{Bodenheimer+2001,GS02}. A number of explanations have been proposed \citep[e.g.][]{Laughlin+11}. For energetic considerations, the most plausible class of models appear to be those which invoke the dissipation of a small fraction (of order 1\% or less) of the stellar flux relatively deep in the planetary interior as a mean to slow the planetary cooling and contraction \citep{GS02,Laine+2008,BS10,ArrasSocrates10}. A definitive test of these models would be an accurate measurement of stellar ages coupled to the determination of precise planetary masses and radii for a statistically significant number of objects. 

Under the assumption that this class of models prevails, it is relatively straightforward to parametrize this 'missing physics' through the dissipation of a fixed fraction of the irradiation energy into the planetary interior. This fraction may be chosen so that model radii for hydrogen-helium planets are always (or almost always) above the observational constraint. The global amount of heavy elements present in the planet may then be inferred from evolution models \citep{Guillot+06}.

\begin{figure}
\includegraphics[width=10cm,angle=0]{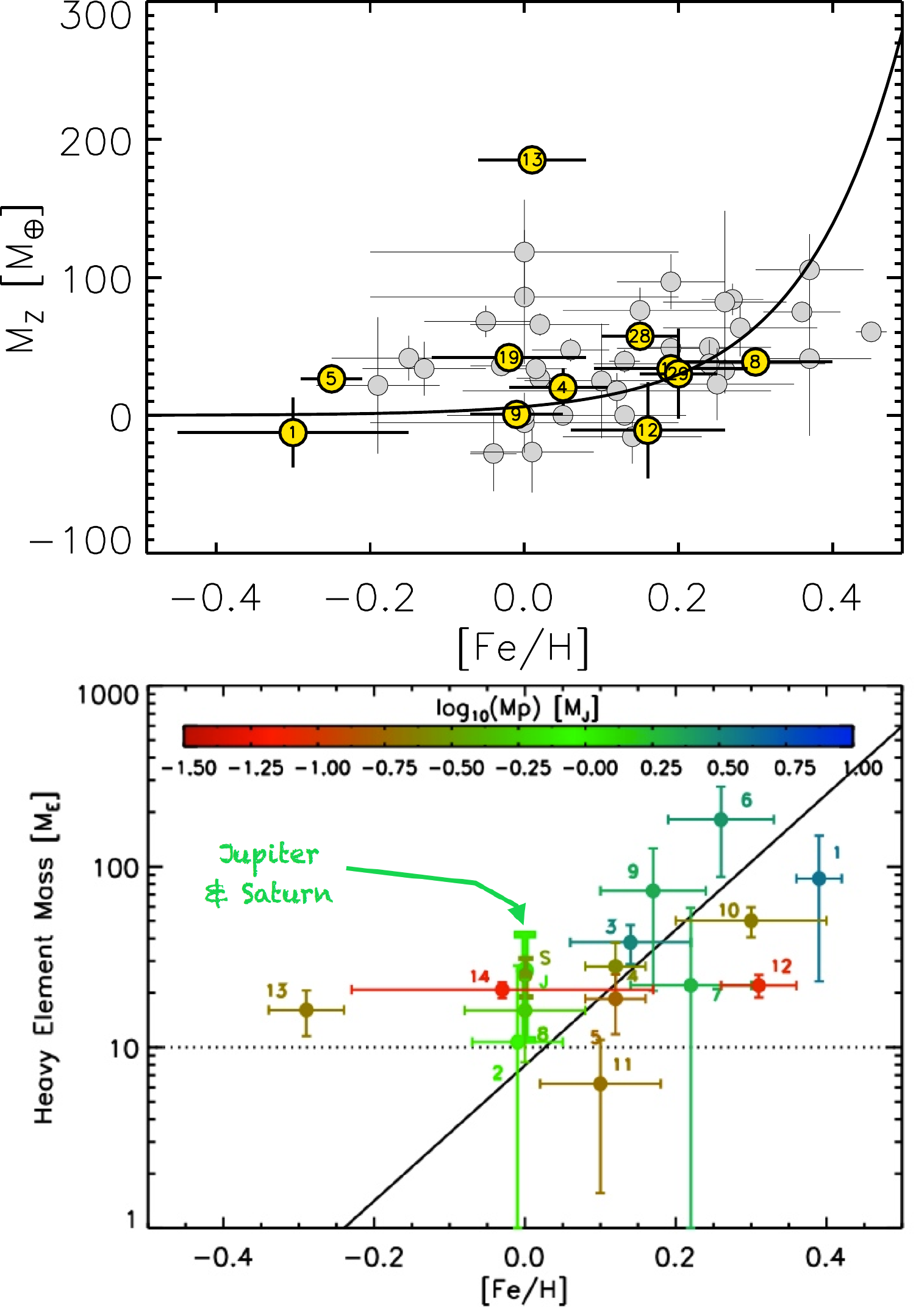}
\caption{Mass fraction of heavy elements in the planets as a function of the metallicity of their parent star. {\it Top panel}: Result for hot Jupiters with masses smaller than $2\,\rm M_{Jup}$. The evolution model assumes that 0.5\% of the incoming irradiation flux is dissipated at the planet's center. Circles which are labeled 1 to 23 correspond to the CoRoT giant planets. Gray symbols correspond to a subset of known transiting systems \cite{Guillot08, Laughlin+11}. Unphysical negative values for $M_Z$ correspond to insufficient heat sources leading to a radius that is larger than observed. [From \cite{Moutou+2013}.] {\it Bottom panel}: Same plot for weakly irradiated planets. [From \cite{MF11}.] In both panels, a curve shows a linear relation between $\log M_Z$ and [Fe/H] fitted to the results for weakly irradiated planets.}
\label{fig:correlation}
\end{figure}

Figure~\ref{fig:correlation} shows the result of such an exercice when applied to hot Jupiters with known masses and radii. In that plot, the mass of heavy elements is plotted against the metallicity of the parent star. For some objects, the chosen (small) value of the dissipation parameter imply an unphysical negative mass of heavy elements. This is for example the case of the still unexplained large radius of CoRoT-2b \citep[see][]{Guillot+Havel2011}. However, two robust results are the requirement of very large masses in heavy elements (above $100\,\rm M_\oplus$) for some objects, and the correlation between the mass of heavy elements inferred in the planets and that in the parent star \citep{Guillot+06,Burrows+07,Guillot08,Laughlin+11,Moutou+2013}. Interestingly, these two results are also observed when one uses a subset of the ensemble of planets that is weakly irradiated and does not require any extra heat source to explain their radius \citep{MF11}. 

The large amount of heavy elements shows that solids were efficiently stored and delivered into the planets. This must have occurred in the circumstellar disk at the time of the formation of the parent star and its planets. Furthermore, the correlation between planet and star 'metallicity' shows that the storage/accretion/delivery of solids was strongly favored by an increased metallicity. This favor scenarios in which giant planets are formed by the accretion of a central core followed by the capture of a hydrogen-helium envelope which may be polluted in variable amounts by dust and planetesimals.

\section{Dynamical interactions between stars and close-in exoplanets}

\begin{figure}
\includegraphics[width=\hsize,angle=0]{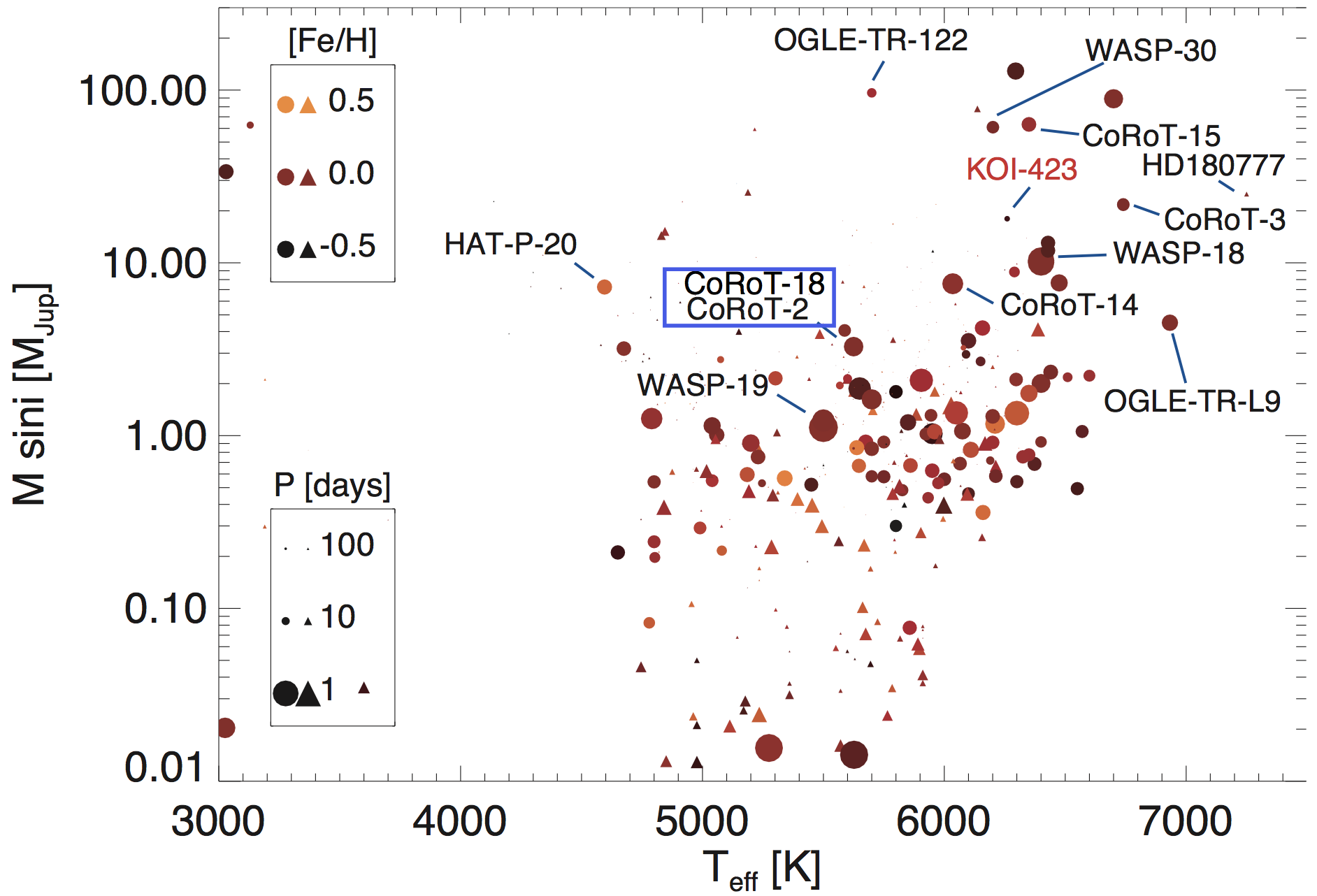}
\caption{Mass (times the sine of the orbital the inclination) of known exoplanets and brown dwarfs as a function of effective temperature of the parent stars. Some of the key systems are labelled. The size of the symbols is inversely proportional to their orbital period. (Large symbols correspond to important tidal interactions between the star and the companion.) [Figure adapted from \cite{Bouchy+2011}.]}
\label{fig:exo_M_Teff}
\end{figure}

Another link between the properties of stars and of their close-in exoplanets is a direct consequence of tidal interactions between them. Figure~\ref{fig:exo_M_Teff} shows the masses of known exoplanets as a function of the effective temperature of their parent star. The symbol sizes are inversely proportional to the planets' orbital periods which effectively helps focusing on short period planets. Obvious observational biases are indicated in the figure: stars with low effective temperature are faint which makes it difficult to observe planets around them with either radial velocimetry or photometry in the visible. Similarly, at high effective temperatures two effects combine against the discovery of small planets: the fast rotation of F-dwarfs makes the spectral lines broader which reduces the sensitivity of radial velocity surveys and the stars become larger which also acts against the discovery of transiting planets by photometry. A paucity of close-in companions is obvious for masses above about $5\,\rm M_{Jup}$ and around stars with effective temperatures between about 5000 and 6000 K. This cannot be explained by an observational bias: if present, massive companions around G-dwarfs would be at least as easily detected as massive companions to F-dwarfs.

Given the small different in mass between F- and G-dwarfs, the different planet populations must be due to a different dynamical evolution caused by different dissipation regimes. This may be shown by defining the inward migration timescale  \citep[see][]{Barker+Ogilvie2009}:
\begin{equation}
\begin{split}
\tau_{\rm mig}=12\,{\rm Ma}&\ \left(Q_*\over 10^6\right)\left(\mp\over 1\,\mjup\right)^{-1}\left(M_*\over 1\,\msol\right)^{8/3}\left(R_*\over 1\,\rsol\right)^{-5}\\
&\times\left(P_{\rm orb}\over 1{\,\rm day}\right)^{13/3}\left(1-{P_{\rm orb}\over P_{\rm spin}}\right)^{-1}
\end{split}
\label{eq:tau_mig}
\end{equation}
where $M_*$ and $R_*$ are the stellar mass and radius, respectively, $\mp$ is the companion's mass, $P_{\rm orb}$ its orbital period and $P_{\rm spin}$ the stellar spin period. $Q'_*\equiv Q_*/k_2$ is the equivalent stellar tidal dissipation factor defined as the ratio between the tidal dissipation factor and the second Love number. The value of $\tau_{\rm mig}$ for known brown dwarfs and exoplanetary companions for an assumed fixed $Q'_*=10^6$ ranges from $10^6$\,yrs and more for G-dwarfs and only about $10^5$\,yrs and more for F-dwarfs (Guillot, Lin \& Morel, in preparation). 

We interpret this difference as arising from the engulfment of massive planets and brown dwarfs around G-dwarfs and their preservation around F-dwarfs. This has two reasons: (i) F-dwarfs are known to be rapid rotators which have a weaker magnetic braking due to their small outer convective zone. They are hence less efficient at extracting angular momentum. On the other hand in systems in which a G-dwarf is spun-up by a close-in companion, stellar winds and magnetic fields yield a rapid loss of angular momentum from the system and a fast runaway migration of the companion onto the central star. (ii) G-dwarfs have a radiative center while F-dwarfs have a central convective zone. This implies that gravity waves propagating in the inner radiative zone may break and dissipate their energy in G-dwarfs but not in F-dwarfs \citep[see][]{Barker+Ogilvie2010}. 

\begin{figure}
\includegraphics[width=\hsize,angle=0]{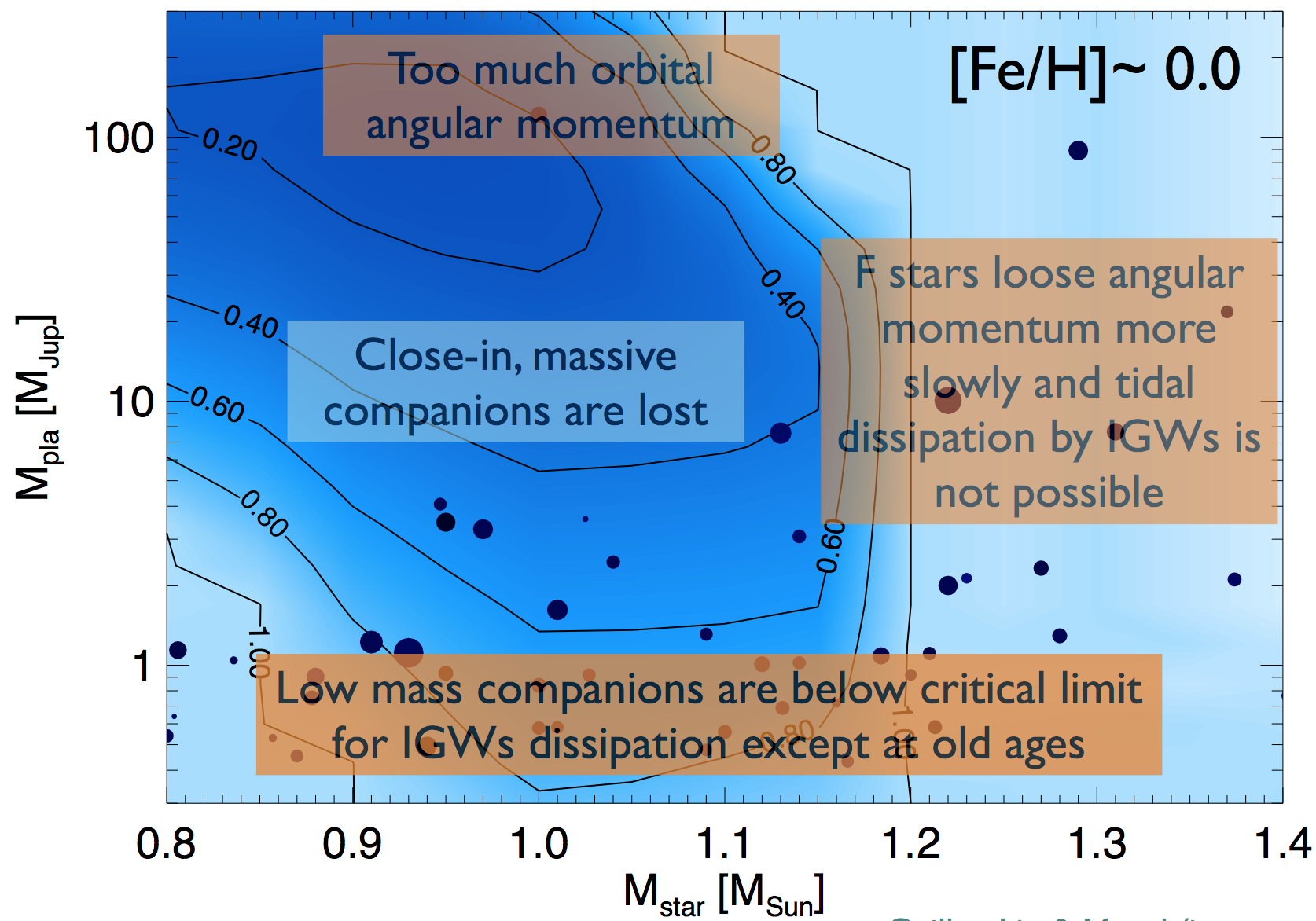}
\caption{Contour plot showing the ratio of the lifetime age to the main sequence lifetime of a companion of mass $M_{\rm pla}$ orbiting around a star of mass $M_{\rm star}$ initially on a 3 day orbit and assuming [Fe/H]=0.0. The model used is the full internal gravity wave model [see text]. The points correspond to the observed low-metallicity population, with $-0.15 \le\rm [Fe/H] < 0.15$. [Figure from Guillot, Lin \& Morel (in preparation).]}
\label{fig:tides}
\end{figure}

Figure~\ref{fig:tides} shows the result of a dynamical model that includes tidal interactions between stars and their companion \citep{Barker+Ogilvie2009}, stellar evolution \citep{Morel+Lebreton2008}, the magnetic braking of stars \citep{Bouvier+1997}, and a consistent calculation of tidal dissipation ($Q'_*$) by gravity waves \citep{Barker+Ogilvie2010}. The latter is calculated consistently by including the evolution of the stellar interior which enter the calculation of $Q'_*$ \citep[see also][]{Barker2011}. The figure shows for stars of 0.8 to 1.4$\,\rm M_\odot$ and companions of 0.2 to 200\,$\rm M_{Jup}$ on an initial 3-day orbital period the fraction of the star's main-sequence lifetime on which the companion is able to survive. 

We find that massive companions around G-dwarfs generate tides that break at the star center, leading to a strong dissipation, their inward migration and eventual demise. The process is much less efficient for small-mass companions essentially because the waves that they exert have a too small amplitude to break except at late ages when the star stars leaving the main sequence. For massive companions, the inward migration is slow because of the large initial angular momentum. Around F-dwarfs, very little migration occurs both because internal gravity waves cannot reach the center and because of their weak magnetic braking. 

Massive close-in planets and brown dwarfs are therefore engulfed preferentially around G-dwarfs in qualitative agreement with the observations. Actually, while this goes in the right direction, it may be argued that the observations show an even stronger deficit of massive companions around G-dwarfs than found by the models. This clearly requires further work.

\section{Conclusion}

A proper understanding of the properties, formation and evolution of planetary systems requires combining observations and theoretical studies of both the planets and their parent stars. This clearly demonstrated by the two examples that we discussed: (i) the determination of the global composition of giant exoplanets which appears to be directly linked to the metallicity of the parent star; (ii) the fate of close-in massive exoplanets and brown dwarfs which depends both on the planetary and stellar properties and the magnitude of their tidal interactions. This is of course not restricted to these two examples. The prospect of discovering thousands of transiting planets and measuring at the same time both the stellar and planetary properties very accurately thanks to a number of ground-based and space-based projects (including CHEOPS, TESS, PLATO) is extremely promising.

%%-----------------------------
%%      your bibliography
%%-----------------------------

\bibliographystyle{aa}
\bibliography{roscoff_guillot+2014}{}

% \begin{thebibliography}{99}
% \bibitem[1994]{alref1} Aalto, S. \etal\  1994, A\&A, 286, 365.
% %%% Using \cite{Bei} in the text
% \bibitem[1986]{Bei} Beichman, C.A., Neugebauer, G., Habing,
%    H., Clegg, P.E. \& Chester, T.C. 1988, editors, {\it ``IRAS Catalogs and
%    Atlases: Explanatory Supplement''}, NASA RP-1190 (Washington: NASA)
% \bibitem[1987]{ref1987} Beichman, C.A. 1987, ARA\&A, 25, 521
% \bibitem[1987]{so1987} Soifer, B.T., Houck, J.R. and Neugebauer, G. 1987, ARAA, 25, 187
% \end{thebibliography}
\end{document}